\pdfoutput=1
\documentclass[conference]{IEEEtran}
\IEEEoverridecommandlockouts

\usepackage{amsmath,amssymb,amsfonts,amsthm,bm}
\usepackage{algorithm}
\usepackage{algorithmic}
\usepackage{balance}
\usepackage[font=footnotesize,skip=1pt]{caption}
\usepackage{cite}
\usepackage{graphicx}
\usepackage{xcolor}

\newtheorem{theorem}{Theorem}

\allowdisplaybreaks

\newcommand\semiHuge{\fontsize{22.72}{27.38}\selectfont}

\begin{document}

\title{\semiHuge Channel Estimation via Successive Denoising in MIMO OFDM Systems: A Reinforcement Learning Approach}

\author{\IEEEauthorblockN{Myeung Suk Oh\IEEEauthorrefmark{1}, Seyyedali Hosseinalipour\IEEEauthorrefmark{1}, Taejoon Kim\IEEEauthorrefmark{2}, Christopher G. Brinton\IEEEauthorrefmark{1} and David J. Love\IEEEauthorrefmark{1}}
\IEEEauthorblockA{\IEEEauthorrefmark{1}Electrical and Computer Engineering, Purdue University, West Lafayette, IN, USA}
\IEEEauthorblockA{\IEEEauthorrefmark{2}Electrical Engineering and Computer Science, University of Kansas, Lawrence, KS, USA}
\IEEEauthorblockA{\IEEEauthorrefmark{1}\{oh223, hosseina, cgb, djlove\}@purdue.edu, \IEEEauthorrefmark{2}taejoonkim@ku.edu}}

\maketitle

\thispagestyle{plain}
\pagestyle{plain}

\begin{abstract}
    In general, reliable communication via multiple-input multiple-output (MIMO) orthogonal frequency division multiplexing (OFDM) requires accurate channel estimation at the receiver. The existing literature largely focuses on denoising methods for channel estimation that depend on either (i)~channel analysis in the time-domain with prior channel knowledge or (ii)~supervised learning techniques which require large pre-labeled datasets for training. To address these limitations, we present a frequency-domain denoising method based on a reinforcement learning framework that does not need a priori channel knowledge and pre-labeled data. Our methodology includes a new successive channel denoising process based on channel curvature computation, for which we obtain a channel curvature magnitude threshold to identify unreliable channel estimates. Based on this process, we formulate the denoising mechanism as a Markov decision process, where we define the actions through a geometry-based channel estimation update, and the reward function based on a policy that reduces mean squared error (MSE). We then resort to Q-learning to update the channel estimates. Numerical results verify that our denoising algorithm can successfully mitigate noise in channel estimates. In particular, our algorithm provides a significant improvement over the practical least squares (LS) estimation method and provides performance that approaches that of the ideal linear minimum mean square error (LMMSE) estimation with perfect knowledge of channel statistics.
\end{abstract}


\vspace{-1mm}
\section{Introduction}\label{sec:introduction}

Many current wireless technologies employ multiple-input multiple-output (MIMO) orthogonal frequency division multiplexing (OFDM) scheme, where multiple antennas and subcarriers are utilized to achieve higher data rates. 
To ensure the robustness of MIMO OFDM, accurate channel estimation is key~\cite{survey1}.
To obtain the channel estimates, it is logical to have the transmitter send a known pilot signal in both the spatial and frequency domains.
The most popular channel estimation criteria based on pilot signals include
linear minimum mean square error (LMMSE) and least squares (LS)~\cite{survey2}.

While LMMSE estimation is optimal in terms of minimizing mean squared error (MSE), it requires prior statistical knowledge, which is not always available in wireless environments. 
LS channel estimation, on the other hand, is a practical lower complexity alternative that can be applied without prior knowledge regarding channel statistics.
However, these benefits come with the cost of performance degradation due to estimation error induced by the noise~\cite{survey2}.

To combat the effect of noise in OFDM LS channel estimation, researchers have proposed various \textit{denoising} techniques~\cite{ce1,ce2,ce3}.
These approaches focus on channel impulse response (CIR) thresholding~\cite{ce1}, significant sample selection~\cite{ce3}, or zero-enforcing on the noise channel subspace~\cite{ce2}, and have proven to be effective in reducing the MSE of LS estimation.
However, all of the prior approaches are channel condition-oriented and are vulnerable to channel dynamics and misalignment to the pre-estimated channel statistics.
Furthermore, these approaches rely on denoising in the time-domain, which increases the computational overhead required to perform a discrete Fourier transform (DFT) per channel realization.

Leveraging machine learning (ML) to re-examine problems has been at the center of wireless communication research recently~\cite{survey4}. 
ML can also be used to denoise LS channel estimates, as demonstrated in~\cite{ML1,ML2,ML3}. 
Gaussian process regression~\cite{ML1} and deep neural networks, called  ChannelNet~\cite{ML2} and ReEsNet~\cite{ML3}, have proven their capabilities refining channel estimation quality substantially.   
These works primarily focus on \textit{supervised learning} techniques, which require training on generally extensive labeled datasets that are acquired from the ideal channel estimation process.
It is unlikely that such labeled training data are always available without exhibiting dependency on noise and spatial and/or temporal channel dynamics commonly found in many 5G mobile use cases~\cite{mobility}.

\textbf{Overview of methodology and contributions:} In this paper, we propose a reinforcement learning (RL)-based channel denoising method to lower the MSE of LS channel estimation in MIMO OFDM systems.
In doing so, we introduce a new successive channel denoising process based on the curvature of channel estimates, and analytically derive the curvature magnitude threshold to identify unreliable estimates among subcarriers.
We then model the denoising process as the problem of finding an optimal sequential order on subcarriers to effectively reduce the MSE of estimation and formulate the denoising as a Markov decision process (MDP).
The actions of the MDP are defined based on a geometric channel estimation update, and the reward function captures the noise reduction obtained through the sequential channel denoising.
To solve the proposed MDP problem, we resort to Q-learning.

Our method eliminates the requirement of genie datasets for training and provides robustness against variation in channel statistics.
Furthermore, our proposed method obtains computational efficiency enhancements by performing denoising in the frequency-domain, eliminating the need for domain conversion.
Our numerical simulations reveal the effectiveness of our method, suggesting a substantial performance gain over LS estimation that approaches the performance of the ideal LMMSE method when perfect channel statistics are available.

\section{System Model}\label{sec:system_model}

In this section, we begin by formalizing MIMO OFDM transmission (Sec.~\ref{ssec:MIMO_OFDM}). 
Then, we introduce the conventional channel estimation methods that we will later use as benchmarks in our analysis (Sec.~\ref{ssec:channel_estimation}).

\subsection{MIMO OFDM Transmission}\label{ssec:MIMO_OFDM}

We consider a MIMO OFDM system with $N_t$ transmit antennas and $N_r$ receive antennas, where each channel path has $L$ CIR taps.
We let $h_{qp}^{(\ell)}$ be the channel of tap $\ell\in\{0,1,\ldots,L-1\}$ between the transmit antenna $p\in\{0,1,\ldots,N_t-1\}$ and the receive antenna $q\in\{0,1,\ldots,N_r-1\}$.
We assume the channel is i.i.d. according to a zero-mean circularly symmetric complex Gaussian with variance $\sigma_\ell^2$, i.e., $h_{qp}^{(\ell)}\sim\mathcal{CN}(0,\sigma_\ell^2)$.
The expected total power $P$ of a channel path is considered to be constant between antennas, i.e., $P = \mathbb{E}\big[\sum^{L-1}_{\ell=0}\vert h_{qp}^{(\ell)}\vert ^2\big] = \sum^{L-1}_{\ell=0}\sigma_\ell^2$, $\forall p,q$.
We assume only $L$ and $P$ are known to the receiver. 
The system employs $K$ subcarriers and a cyclic prefix of length $L-1$.

The frequency-domain input-output relationship for subcarrier $k\in\{0,\ldots,K-1\}$ of an OFDM symbol is given by
\begin{equation}
    \textbf{y}[k]=\textbf{H}[k]\textbf{x}[k]+\textbf{w}[k],
    \label{eq:relationship}
\end{equation}
where $\textbf{y}[k]=[y_{0}[k],y_{1}[k],\ldots,y_{N_r-1}[k]]^T\in\mathbb{C}^{N_r}$ and $\textbf{x}[k]=[x_{0}[k],x_{1}[k],\ldots,x_{N_t-1}[k]]^T\in\mathbb{C}^{N_t}$ are the $k$th subcarrier frequency-domain receive and transmit symbol vectors, respectively.
The transmit symbols are assumed to be unit power, i.e., $\mathbb{E}[\vert x_{p}[k]\vert ^2]=1$, $\forall p$.
In~\eqref{eq:relationship}, $\textbf{w}[k]=[w_{0}[k],w_{1}[k],\ldots,w_{N_r-1}[k]]^T\in\mathbb{C}^{N_r}$ is the noise vector with entries i.i.d. according to $w_{q}[k] \sim \mathcal{CN}(0,\sigma_w^2)$, $\forall q$, and $\textbf{H}[k] = [H_{qp}^{(k)}]\in\mathbb{C}^{N_r\times N_t}$ denotes the MIMO channel matrix of subcarrier $k$ where
\begin{equation}
    H_{qp}^{(k)}=\sum^{L-1}_{\ell=0}h_{qp}^{(\ell)}e^{-j2\pi \ell\frac{k}{K}}.
    \label{eq:H_k}
\end{equation}
To obtain the equalized symbol vector for subcarrier $k$, denoted by $\tilde{\textbf{x}}[k]=[\tilde{x}_{0}[k],\tilde{x}_{1}[k],\ldots,\tilde{x}_{N_t-1}[k]]^T\in\mathbb{C}^{N_t}$, a zero-forcing equalizer is applied to each $\textbf{y}[k]$ as
\begin{equation}
    \tilde{\textbf{x}}[k]=\textbf{H}[k]^{H}(\textbf{H}[k]\textbf{H}[k]^{H})^{-1}\textbf{y}[k],
    \label{eq:ZF}
\end{equation}
where $(\cdot)^H$ refers to the conjugate transpose. 
In~\eqref{eq:ZF}, we assume the case where $N_t\geq N_r$.

We consider a frame-based transmission scenario where each frame consists of a single pilot signal for channel estimation and $D$ data signals for data transfer.
We also assume the channel to be block-fading, where the channel is constant over the duration of $D+1$ OFDM symbols and varies across frames. 
The system aims to estimate the channel from the pilot signal to correctly detect data symbols within the same frame.

\subsection{Channel Estimation}\label{ssec:channel_estimation}

We consider two representative channel estimation approaches: LS and LMMSE. 
\subsubsection{LS} Suppose each transmit antenna sends its pilot symbol vector denoted by $\widehat{\textbf{x}}_{p}=[\widehat{x}_{p}[0],\widehat{x}_{p}[1],\ldots,\widehat{x}_{p}[K-1]]^T\in\mathbb{C}^{K}$ at $N_t$ different times to avoid interference. 
Given the pilot observation $\widehat{\textbf{y}}_{q}=[\widehat{y}_{q}[0],\widehat{y}_{q}[1],\ldots,\widehat{y}_{q}[K-1]]^T\in\mathbb{C}^{K}$ at the $q$th receive antenna, the LS channel estimate denoted by $\overline{\textbf{H}}^{\text{LS}}_{qp}=[\widehat{H}_{qp}^{(0)},\widehat{H}_{qp}^{(1)},\ldots,\widehat{H}_{qp}^{(K-1)}]^T\in\mathbb{C}^{K}$ is obtained as follows: 
\begin{equation}
    \overline{\textbf{H}}^{\text{LS}}_{qp}=\text{diag}(\widehat{\textbf{x}}_{p})^{-1}\widehat{\textbf{y}}_{q}=\overline{\textbf{H}}_{qp}+\text{diag}(\widehat{\textbf{x}}_{p})^{-1}\textbf{w}_{q},
    \label{eq:LS}
\end{equation}
where $\overline{\textbf{H}}_{qp}=[H_{qp}^{(0)},H_{qp}^{(1)},\ldots,H_{qp}^{(K-1)}]^T\in\mathbb{C}^{K}$ and $\textbf{w}_{q}=[w_{q}[0],w_{q}[1],\ldots,w_{q}[K-1]]^T\in\mathbb{C}^{K}$ are the true channel vector between the corresponding transmit and receive antenna and the noise vector at the receiver, respectively~\cite{survey2}. 
The expression in~\eqref{eq:LS} can be equivalently written for the $k$th subcarrier as
\begin{align}
    \widehat{H}_{qp}^{(k)}&=\frac{\widehat{y}_{q}[k]}{\widehat{x}_{p}[k]}=H_{qp}^{(k)}+\frac{w_{q}[k]}{\widehat{x}_{p}[k]},
    \label{eq:freq}
\end{align}
which contains both the true channel and the noise.

\subsubsection{LMMSE} Provided the LS estimate in~\eqref{eq:LS}, the LMMSE channel estimate $\overline{\textbf{H}}^{\text{LMMSE}}_{qp}\in\mathbb{C}^{K}$ can be succinctly written as  
\begin{align}
    \overline{\textbf{H}}^{\text{LMMSE}}_{qp}&=\textbf{R}_{qp}\left(\textbf{R}_{qp}+\frac{\sigma^2_w}{\mathbb{E}\left[\vert x_{p}[k]\vert ^2\right]}\textbf{I}_K\right)^{-1}\overline{\textbf{H}}^{\text{LS}}_{qp},
    \label{eq:MMSE}
\end{align}
where $\textbf{R}_{qp}=\mathbb{E}\left[\overline{\textbf{H}}_{qp}\overline{\textbf{H}}_{qp}^H\right]\in\mathbb{C}^{K\times K}$ is the correlation matrix of the channel vector $\overline{\textbf{H}}_{qp}$ and $\textbf{I}_K$ is the $K\times K$ identity matrix~\cite{survey2}.
The implication of~\eqref{eq:MMSE} is that a priori channel statistics must be known to compute~\eqref{eq:MMSE}, which are not always available in practical wireless networks~\cite{survey1}, making this solution unrealistic. 
This motivates the proposed learning-based methodology presented the next section.  

\section{Proposed Learning-based Methodology}\label{sec:proposed}

\subsection{Rationale of Approach}\label{ssec:rationale}

With the assumption that $L\ll K$ -- which is valid in many OFDM systems~\cite{survey2,ce1,ce3} -- the channel $\mathbf{H}[k]$ in~\eqref{eq:relationship} will change slowly across subcarriers while the uncorrelated noise $\mathbf{w}[k]$ will vary rapidly.
Although it is difficult to obtain accurate information on the correlation between channels in the presence of noise, the channel estimation can still reveal information about the expected behavior of adjacent subcarriers.
We seek to exploit this information to determine whether our estimates are reliable (i.e., whether the estimation has been severely corrupted by the noise) and denoise them if needed.
Specifically, we will develop a channel denoising method in which the estimations from adjacent subcarriers are jointly used to conduct sequential denoising in subcarriers, where the initial estimate is obtained via LS estimation.

As the first step, we introduce channel curvature to capture the degree of noise contamination, and obtain the threshold on the channel curvature magnitude that differentiates between reliable and unreliable channel estimates (Sec.~\ref{ssec:threshold}).
Then, we introduce a successive subcarrier denoising method and formulate it as an MDP, for which Q-learning is applied to find optimal denoising decisions. (Sec.~\ref{ssec:denoising}).

\subsection{Channel Curvature and Denoising Threshold}\label{ssec:threshold}

Suppose our system acquires an LS-estimated channel vector $\overline{\textbf{H}}^{\text{LS}}_{qp}$ and we want to obtain the relationship between each $\widehat{H}_{qp}^{(k)}$ and its adjacent subcarriers.
The first-order gradient is a natural candidate, as the regression slope can quantify the relative position of data with respect to the neighboring points.
However, since the regression slope is defined as a sum of multiple weighted slopes~\cite{slope}, the issue of weight adjustment arises, making the gradient an ineffective approach for capturing the relationship.
On the other hand, the \textit{curvature}, i.e., the second-order gradient, consistently reflects the relationship between $\widehat{H}_{qp}^{(k)}$ and its adjacent channels. 
This motivates us to propose the curvature of ${\widehat{H}}_{qp}^{(k)}$ as a measure of its reliability.

From the estimated channel vector $\overline{\textbf{H}}^{\text{LS}}_{qp}$, we approximate the curvature of each $\widehat{H}_{qp}^{(k)}$, denoted by $\widehat{C}_{qp}^{(k)}$, as follows:
\begin{align}
    \widehat{C}_{qp}^{(k)} &= \left(\widehat{H}_{qp}^{(k+1)}-\widehat{H}_{qp}^{(k)}\right)-\left(\widehat{H}_{qp}^{(k)}-\widehat{H}_{qp}^{(k-1)}\right) \nonumber \\
    &= \widehat{H}_{qp}^{(k+1)}-2\widehat{H}_{qp}^{(k)}+\widehat{H}_{qp}^{(k-1)}.
    \label{eq:curv_estimate}
\end{align}
Note that for the cases of $k=0$ and $k=K-1$, we impose the circular shift property to have $\widehat{H}_{qp}^{(k-1)} = \widehat{H}_{qp}^{(K-1)}$ for $k=0$ and $\widehat{H}_{qp}^{(k+1)} = \widehat{H}_{qp}^{(0)}$ for $k=K-1$.

We next aim to obtain the \textit{curvature magnitude threshold} $\widetilde{C}$ that classifies unreliable channel estimates.
To find this threshold, we first obtain the curvature of actual channel between transmit antenna $p$ and receive antenna $q$ for subcarrier $k$, denoted by ${C}^{(k)}_{qp}$, based on the second derivative of~\eqref{eq:H_k}:
\begin{equation}
    C^{(k)}_{qp} = \frac{d^2H_{qp}^{(k)}}{d k^2}=\sum^{L-1}_{\ell=1}-\left(\frac{2\pi \ell}{K}\right)^2h^{(\ell)}_{qp}e^{-j2\pi \ell\frac{k}{K}}.
    \label{eq:C_k}
\end{equation}
Since the values of $\{h^{(\ell)}_{qp}\}^{L-1}_{\ell=1}$ randomly change over every transmission frame, the value of $C_{qp}^{(k)}$ is also random and time-varying.
From~\eqref{eq:C_k}, we derive an upper bound on the expected magnitude of $C^{(k)}_{qp}$ in the following theorem:
\begin{theorem}
For an $N_t\times N_r$ MIMO OFDM $L$-tap channel with channel power $P$, the upper bound on the expected magnitude of $C^{(k)}_{qp}$ is given by
\begin{equation}
    \hspace{-2mm}\mathbb{E}\left[\big\vert C^{(k)}_{qp}\vert \right]\leq\left(\frac{2\pi}{K}\right)^2\hspace{-1mm}\xi(1,2)\sqrt{(P-\sigma^2_{0})\hspace{-1mm}\sum^{L-1}_{\ell=1}\ell^4}\triangleq \Bar{C}(\sigma^2_0),
    \label{eq:theorem1}
\end{equation}
where $\xi(x,y)=\sqrt{2 x \log{2y}}$.
\begin{proof}
We first derive a simple upper bound on the expected magnitude of curvature using~\eqref{eq:C_k}:
\begin{align}
    \mathbb{E}\left[\big\vert C^{(k)}_{qp}\big\vert \right] &=\mathbb{E}\left[\bigg\vert \sum^{L-1}_{\ell=1}-\left(\frac{2\pi \ell}{K}\right)^2h^{(\ell)}_{qp}e^{-j2\pi \ell\frac{k}{K}}\bigg\vert \right] \nonumber \\
    &\leq \mathbb{E}\left[\sum^{L-1}_{\ell=1}\bigg\vert -\left(\frac{2\pi \ell}{K}\right)^2h^{(\ell)}_{qp}e^{-j2\pi \ell\frac{k}{K}}\bigg\vert \right] \nonumber \\
    &=\sum^{L-1}_{\ell=1}\left(\frac{2\pi \ell}{K}\right)^2\mathbb{E}\left[\big\vert h^{(\ell)}_{qp}\big\vert \right],
    \label{1}
\end{align}
where the inequality holds from the triangle inequality, and the equality holds with the expectation directly applied to $\vert h^{(\ell)}_{qp}\vert $.

For a sequence of Gaussian random variables $X_1,\ldots,X_n$ where $X_i\sim\mathcal{N}(0,\sigma^2)$, $\forall i$, the following holds~\cite{maxGauss}:
\begin{equation}
    \mathbb{E}\left[\underset{i\in\{1,\ldots,n\}}{\text{max}}\vert X_i\vert \right]\leq\sqrt{2\sigma^2\log{2n}}\triangleq \xi(\sigma^2,n).
    \label{eq:maxGaussian}
\end{equation}\vspace{-2mm}

Using~\eqref{eq:maxGaussian}, the expectation of $\vert h^{(\ell)}_{qp}\vert $ in~\eqref{1} can be upper bounded as follows:
\begin{align}
    &\hspace{-1.5mm}\mathbb{E}\left[\big\vert h^{(\ell)}_{qp}\big\vert \right]=\mathbb{E}\left[\sqrt{(\Re\{h^{(\ell)}_{qp}\})^2+(\Im\{h^{(\ell)}_{qp}\})^2}\right]\nonumber\\
    &\leq\mathbb{E}\hspace{-0.5mm}\left[\sqrt{2\big(\text{max} \big\{ \vert \Re\{h^{(\ell)}_{qp}\}\vert , \vert \Im\{h^{(\ell)}_{qp}\}\vert \big\}\big)^2}\right]\hspace{-0.5mm}\leq\sigma_\ell \xi(1,2).\hspace{-1mm}
    \label{2}
\end{align}
The equality is from the definition of complex Gaussian random variable $h^{(\ell)}_{qp} = \Re\{h^{(\ell)}_{qp}\} + j \Im\{h^{(\ell)}_{qp}\}$, where $\Re\{h^{(\ell)}_{qp}\}, \Im\{h^{(\ell)}_{qp}\} \sim \mathcal{N}(0,\frac{\sigma^2_\ell}{2})$.
The last inequality holds from applying~\eqref{eq:maxGaussian}, which yields $\mathbb{E}\Big[\text{max} \big\{\vert  \Re\{h^{(\ell)}_{qp}\}\vert , \vert \Im\{h^{(\ell)}_{qp}\}\vert  \big\} \Big] \leq \xi(\sigma^2_\ell/2,2)$, and by noting that $\xi(x^2,y)=x\cdot \xi(1,y)$.

Applying~\eqref{2} to~\eqref{1}, we get
\begin{align}
    \mathbb{E}\left[\big\vert C^{(k)}_{qp}\big\vert \right] &\leq \sum^{L-1}_{\ell=1}\left(\frac{2\pi \ell}{K}\right)^2\sigma_\ell \xi(1,2) \nonumber \\
    &\leq\left(\frac{2\pi}{K}\right)^2\xi(1,2)\sqrt{\sum^{L-1}_{\ell=1}\ell^4\cdot\sum^{L-1}_{\ell=1}\sigma_\ell^2} \nonumber \\
    &=\left(\frac{2\pi}{K}\right)^2\xi(1,2)\sqrt{(P-\sigma^2_0)\sum^{L-1}_{\ell=1}\ell^4}.
\end{align}
The inequality in the second line is obtained via the Cauchy-Schwarz inequality of
$\sum\ell^2\sigma_\ell\leq\sqrt{\sum(\ell^2)^2}\cdot\sqrt{\sum(\sigma_\ell)^2}$, and the equality holds since $P=\sum^{L-1}_{\ell=0}\sigma_\ell^2$.
\end{proof}
\label{theorem}
\end{theorem}
\vspace{-3mm}
\textit{Remark:} Since $\Bar{C}(\sigma^2_0)$ in~\eqref{eq:theorem1} is the maximum magnitude of subcarrier channel curvature expected from a MIMO OFDM $L$-tap channel with channel power $P$, we want to have $\widetilde{C}=\Bar{C}(\sigma^2_0)$.
However, obtaining $\Bar{C}(\sigma^2_0)$ requires the knowledge on $\sigma^2_0$, which is not the case we can consider.
We therefore introduce the term $\widehat{\sigma}^2_{0}=\frac{1}{N_tN_r}\sum^{N_t-1}_{p=0}\sum^{N_r-1}_{q=0}\vert \frac{1}{K}\sum^{K-1}_{k=0}\widehat{H}_{qp}^{(k)}\vert ^2$ to approximate $\sigma_0^2$.
We point to the DFT operation in~\eqref{eq:H_k}, which in the large $K$ regime gives $\frac{1}{K}\sum^{K-1}_{k=0}\widehat{H}_{qp}^{(k)}\approx \mathbb{E}[\widehat{H}_{qp}^{(k)}]=\mathbb{E}[H_{qp}^{(k)}]+\mathbb{E}[w_q^{(k)}]=h^{(0)}_{qp}$. 
If the average of $\vert \frac{1}{K}\sum^{K-1}_{k=0}\widehat{H}_{qp}^{(k)}\vert ^2\approx \vert h_{qp}^{(0)}\vert ^2$ is taken over $N_tN_r$ channel links, we obtain $\widehat{\sigma}^2_0$ that approximates $\sigma_0^2$.
We can now evaluate $\Bar{C}(\widehat{\sigma}^2_0)$ to approximate $\Bar{C}(\sigma^2_0)$ and set $\widetilde{C}=\Bar{C}(\widehat{\sigma}^2_0)$.

For our denoising, we classify the estimated channel $\widehat{H}_{qp}^{(k)}$ as reliable if its curvature satisfies
\begin{equation}
    \vert \widehat{C}^{(k)}_{qp}\vert \leq \widetilde{C},
    \label{eq:curv_ineq}
\end{equation}
and consider $\widehat{H}_{qp}^{(k)}$ as unreliable otherwise. 

\subsection{Successive Denoising Formulation and Optimization}\label{ssec:denoising}

\subsubsection{MDP denoising formulation} We aim to make the best sequential decisions on which subcarrier to select and denoise.
Suppose we initially \emph{observe} $M$ channel estimates as an $M$-dimensional state $\textbf{S}$, and take an action $a$ to \emph{denoise} a single channel estimate that fails to satisfy~\eqref{eq:curv_ineq}.
Once the action $a$ is taken, a different set of $M$ channel estimates, denoted $\textbf{S}'$, will be observed.
We then consider $\textbf{S}'$ as our new state and take another action $a'$ to perform denoising.
If we repeat this observe-and-denoise process until it reaches a terminating state where there is no subcarrier to denoise, our denoising problem can be formulated as an MDP~\cite{RL}.

\textbf{State}: Formally, we define the state as a set of channel estimates:
\begin{equation}
    \textbf{S}(i)=\big[f_Q(\widehat{H}_{qp}^{(i)}), f_Q(\widehat{H}_{qp}^{(i+1)}), \ldots, f_Q(\widehat{H}_{qp}^{(i+M-1)})\big],
    \label{eq:state}
\end{equation}
where $i\in\{0,1,\ldots,K-M\}$ indicates a subcarrier index from which the $M$-dimensional state is obtained out of $K$ subcarriers, and $f_Q(x)$ is a quantization function given by
\begin{equation}
    f_Q(x) = \Delta\cdot\left\lfloor\frac{\Re\{x\}}{\Delta}+\frac{1}{2}\right\rfloor + j\left(\Delta\cdot\left\lfloor\frac{\Im\{x\}}{\Delta}+\frac{1}{2}\right\rfloor\right),
\end{equation}
with quantization step size $\Delta$. This quantization process allows us to represent the environment observations with a finite number of states~\cite{MDP}. Using~\eqref{eq:state}, for an arbitrary value of $i$, the quantized channel estimates from the $i$th to $i+M-1$th subcarriers form an $M$-dimensional state.

\textbf{Action}: The action in our problem is an index indicating which channel estimate to denoise.
From a given state $\textbf{S}(i)$, a set of possible actions $\mathcal{A}$ is formed as follows:
\begin{equation}
    \mathcal{A}(i) =\Big\{a\in\{0,1,\ldots,M-1\}\;:\;\vert \widehat{C}^{(i+a)}_{qp}\vert >\widetilde{C}\Big\}.
\end{equation}
For selecting an action from $\mathcal{A}$, any decision-making strategy that leads to a policy improvement can be used; a common choice is $\epsilon$-greedy~\cite{RL}, which we adopt in this paper.

Once an action $a\in\mathcal{A}(i)$ is chosen, the next state $\textbf{S}'(i)$ is observed through the transition function $T(\textbf{S}(i),a)$ defined as
\begin{align}
    \hspace{-4mm}\textbf{S}'(i)&=T\big(\textbf{S}(i),a\big) \nonumber \\
    &=\big[f_Q(\widehat{H}_{qp}^{\prime(i)}),f_Q(\widehat{H}_{qp}^{\prime(i+1)}), \ldots, f_Q(\widehat{H}_{qp}^{\prime(i+M-1)})\big],\hspace{-4mm}
    \label{eq:transition}
\end{align}
where we propose to update the channel estimates using the following criterion for each $m\in\{0,1,\ldots,M-1\}$:
\begin{align}
    \hspace{-0.3cm}\widehat{H}&_{qp}^{\prime(i+m)} \!=\!
    \begin{cases}
        Z_{qp}^{(i+m)}\!+\!\frac{\widetilde{C}}{2}\cdot\frac{\widehat{H}_{qp}^{(i+m)}-Z_{qp}^{(i+m)}}{\vert \widehat{H}_{qp}^{(i+m)}-Z_{qp}^{(i+m)}\vert }\!\!\! &\text{if }\;\! m=a\\
        \widehat{H}_{qp}^{(i+m)} \!\!\! &\text{otherwise}
    \end{cases}
    \label{update}
\end{align}
with $Z_{qp}^{(x)}=(\widehat{H}_{qp}^{(x-1)}+\widehat{H}_{qp}^{(x+1)})/2$.
The reasoning for this estimation update is as follows. 
Substituting $\widehat{C}^{(k)}_{qp}$ in~\eqref{eq:curv_ineq} with the definition in~\eqref{eq:curv_estimate} yields
\begin{equation}
    \vert \widehat{H}_{qp}^{(k+1)}-2\widehat{H}_{qp}^{(k)}+\widehat{H}_{qp}^{(k-1)}\vert ^2\leq \widetilde{C}^2.
    \label{eq:111}
\end{equation}
Then, the above inequality can be expressed as a circle as follows:
\begin{equation}
    \left(\Re\{\widehat{H}_{qp}^{(k)}-Z_{qp}^{(k)}\}\right)^2+\left(\Im\{\widehat{H}_{qp}^{(k)}-Z_{qp}^{(k)}\}\right)^2\leq\frac{\widetilde{C}^2}{4}.
    \label{eq:222}
\end{equation}
Given two channel estimates $\widehat{H}_{qp}^{(k-1)}$ and $\widehat{H}_{qp}^{(k+1)}$, $\widehat{H}_{qp}^{(k)}$ must be located within a circle centered at $Z_{qp}^{(k)}$ with radius $\widetilde{C}/2$ to satisfy~\eqref{eq:curv_ineq}. 
The estimation update given by~\eqref{update} corresponds to the minimal displacement such that the updated point $\widehat{H}^{\prime(k)}_{qp}$ is located on the circle described in~\eqref{eq:222}.

\textbf{Reward}: Once $\textbf{S}'(i)$ is observed, the reward is obtained based on the effectiveness of the action taken in terms of the problem objective.
For minimizing the MSE of our channel estimation, we use the following expression for the reward:
\vspace{-0.5mm}
\begin{equation}
    \vspace{-0.5mm}
    r(\textbf{S}(i),a)\hspace{-0.5mm}= \hspace{-0.5mm}\frac{1}{K}\hspace{-0.5mm}\sum^{K-1}_{k=0}\hspace{-1mm}\left(\big\vert \widehat{H}_{qp}^{(k)}-\widehat{h}^{(0)}_{qp}\big\vert ^2\hspace{-1.5mm}-\hspace{-0.5mm}\big\vert \widehat{H}^{\prime(k)}_{qp}-\widehat{h}^{(0)}_{qp}\big\vert ^2\right),
    \label{eq:reward}
\end{equation}
where $\widehat{h}^{(0)}_{qp}=\frac{1}{K}\sum_{k=0}^{K-1}\widehat{H}_{qp}^{(k)}$.
This reward function is the change in variance of channel estimates along subcarriers upon taking an action.
For large $K$, by the law of large numbers,~\eqref{eq:reward} can be written as:
\begin{align}
    &\hspace{-0.4cm} \mathbb{E}\Big[\big\vert \widehat{H}_{qp}^{(k)} -\mathbb{E}[\widehat{H}_{qp}^{(k)}]\big\vert ^2\Big]-\mathbb{E}\Big[\big\vert \widehat{H}_{qp}^{\prime(k)}-\mathbb{E}[\widehat{H}_{qp}^{(k)}]\big\vert ^2\Big] \nonumber \\
    &= \mathbb{E}\Big[\big\vert \widehat{H}_{qp}^{(k)}-h^{(0)}_{qp}\big\vert ^2\Big]-\mathbb{E}\Big[\big\vert \widehat{H}^{\prime(k)}_{qp}-h^{(0)}_{qp}\big\vert ^2\Big] \nonumber \\
    &=(P+\sigma^2_w-\sigma^2_0)-(P+\sigma^2_{w'}-\sigma^2_{0}) = \sigma^2_w-\sigma^2_{w'},
    \label{eq:reward2}
\end{align}
where $\sigma^2_{w'}$ is the remaining noise variance after taking the action $a$. In~\eqref{eq:reward2}, the first equality holds since $\mathbb{E}[\widehat{H}^{(k)}]=\mathbb{E}[H^{(k)}]+\mathbb{E}[w[k]]=h^{(0)}$, and the second equality holds from our assumption on uncorrelated channels and noise.
Thus, a greater reward is attributed to an action that eliminates more noise.
Since the MSE of LS channel estimation is proportional to the noise variance~\cite{ce1}, our reward effectively captures and reflects the improvement in MSE upon taking the action $a$.

\subsubsection{Q-learning-based solution} Considering our MDP-based denoising, the sequential order in which channel estimates are selected and denoised becomes an important factor, especially with a low signal-to-noise ratio (SNR) condition where multiple consecutive subcarriers are likely to be unreliable.
In the MDP we consider, $\textbf{S}'$ from any state-action pair $(\mathbf{S}, a)$ is deterministic (i.e., $P(\textbf{S}'\vert \textbf{S},a)=1$).
It is hence possible to apply a brute force search or SARSA learning~\cite{RL} over all combinations of denoising orders, but this will impose a significant amount of computational overhead.

Instead, to learn the optimal sequential denoising order, we adopt Q-learning~\cite{RL}, which seeks to learn the quality of actions while maximizing the cumulative reward.
Unlike supervised learning algorithms, it does not require a training stage as its learning is executed through exploration and exploitation steps.
Q-learning will find the optimal policy for any finite MDP (i.e., with finite state and action spaces)~\cite{RL}, as is the case in our setting.

Using the MDP parameters we established, the state-action quality $Q(\textbf{S}(i),a)$ of Q-learning is updated using the following value iteration~\cite{RL}:
\vspace{-0.5mm}
\begin{align}
    \vspace{-0.5mm}
    \hspace{-1mm}Q&(\textbf{S}(i),a) \nonumber \leftarrow Q(\textbf{S}(i),a) \\& +\alpha(r(\textbf{S}(i),a)+\gamma\underset{a'}{\text{ max }}Q(\textbf{S}'(i),a')-Q(\textbf{S}(i),a))
    \label{eq:Q},
\end{align}
where $\alpha$ and $\gamma$ are the learning rate and the discount factor, respectively.
The Bellman update in~\eqref{eq:Q} allows the current state-action pair to consider its potential future states and actions.
In our context, this update performs successive subcarrier denoising leading to the maximum noise reduction.

\vspace{-1mm}
\subsection{Additional Optimization via Threshold Update}\label{ssec:threshold_update}

We also introduce a feedback scheme that further adjusts the threshold $\widetilde{C}$ to improve the overall denoising performance.
This allows our algorithm to evaluate the effectiveness of $\widetilde{C}$ on current channel estimates and improve its future denoising.
We define the cumulative feedback $F$ to be updated after each complete procedure of denoising as $F \leftarrow F + \Delta F$, where $\Delta F$ is the variance of the remaining noise given by
\vspace{-3mm}

{\small
\begin{align}
    \hspace{-4mm}\Delta F \hspace{-.5mm}=\hspace{-.5mm} \frac{1}{N_tN_rK}\hspace{-1mm}\sum^{N_t-1}_{p=0}\sum^{N_r-1}_{q=0}\sum^{K-1}_{k=0}\hspace{-0.5mm}\big\vert \widehat{H}_{qp}^{(k)}\big\vert ^2\hspace{-1mm}-\hspace{-.5mm}P\label{feedback}\hspace{-.1mm} \approx\hspace{-.1mm}  \mathbb{E}\hspace{-0.5mm}\left[\big\vert \widehat{H}_{qp}^{(k)}\big\vert ^2\right]\hspace{-0.8mm}-\hspace{-.5mm}P.\hspace{-4mm}
\end{align}
}
\vspace{-3mm}

In the next denoising procedure, the curvature threshold is updated as follows:
\vspace{-2mm}
\begin{equation}
    \widetilde{C} = \Bar{C}(\widehat{\sigma}^2_0) - \Big(\frac{2\pi}{K}\Big)^2F.
    \label{eq:F_update}
\end{equation}
The scaling term $(\frac{2\pi}{K})^2$ is from~\eqref{eq:C_k}, reflecting the impact of noise on the channel curvature.

The overall denoising algorithm developed in this section is summarized in Algorithm 1.

\vspace{-2mm}
\begin{algorithm}[h]
{\small
\caption{Learning-based successive denoising algorithm.}
\begin{algorithmic}
\FOR {each frame received}
\STATE Acquire $\overline{\textbf{H}}^{\text{LS}}_{qp}$ for all transmitters $p$ and receivers $q$
\STATE Acquire and adjust $\widetilde{C}$ using~\eqref{eq:theorem1} and~\eqref{eq:F_update}
\FOR {every $(q,p)$ pair}
\WHILE {$\vert \widehat{C}^{(k)}_{qp}\vert > \widetilde{C}$ for any $k$}
\STATE Select random subcarrier $k$ from $\{0,\ldots,K-M\}$
\STATE Initialize state $\textbf{S}(k)$
\WHILE {$\mathcal{A}(k) \neq \phi$}
\STATE Select action $a$ from $\mathcal{A}(k)$ using $\epsilon$-greedy
\STATE Observe $\textbf{S}'(k)$ using~\eqref{eq:transition} and~\eqref{update}
\STATE Compute reward $r(\textbf{S}(i),a)$ using~\eqref{eq:reward}
\STATE Update quality $Q(\textbf{S}(k),a)$ using~\eqref{eq:Q}
\STATE Update state $\textbf{S}(k)\leftarrow\textbf{S}'(k)$
\ENDWHILE
\ENDWHILE
\ENDFOR
\STATE Compute $\Delta F$ using~\eqref{feedback}
\STATE Update $F\leftarrow F+\Delta F$
\ENDFOR
\end{algorithmic}
}
\end{algorithm}

\vspace{-3mm}
\section{Numerical Results and Discussion}\label{sec:numerical}

\begin{figure*}[ht]
    \centering
    \minipage{0.32\textwidth}
        \includegraphics[width=\linewidth]{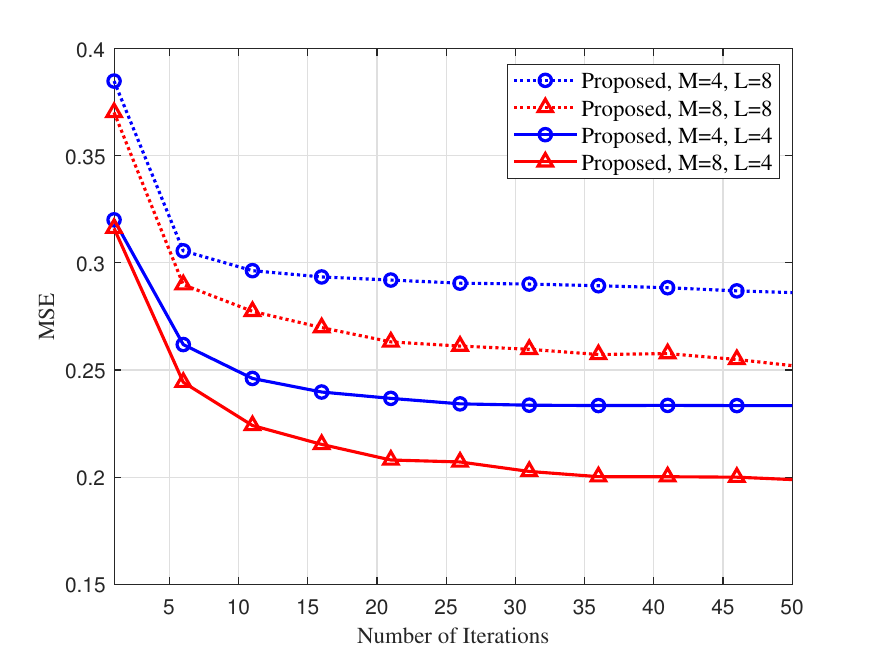}
        \caption{MSE of our method vs. number of learning iterations for different values of state dimension $M$ and channel length $L$ under fixed channel sets.}
        \label{fig:Iter}
    \endminipage\hfill
    \minipage{0.32\textwidth}
        \includegraphics[width=\linewidth]{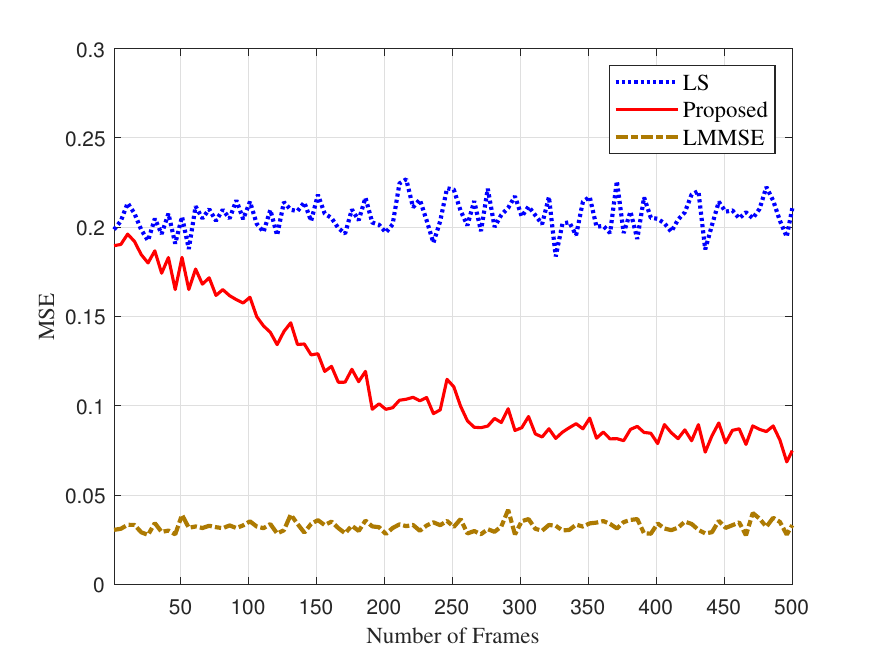}
        \caption{MSE comparison between our method, LS, and LMMSE vs. the number of transmitted frames, where channels are i.i.d. generated.}
        \label{fig:Frames}
    \endminipage\hfill
    \minipage{0.32\textwidth}
        \includegraphics[width=\linewidth]{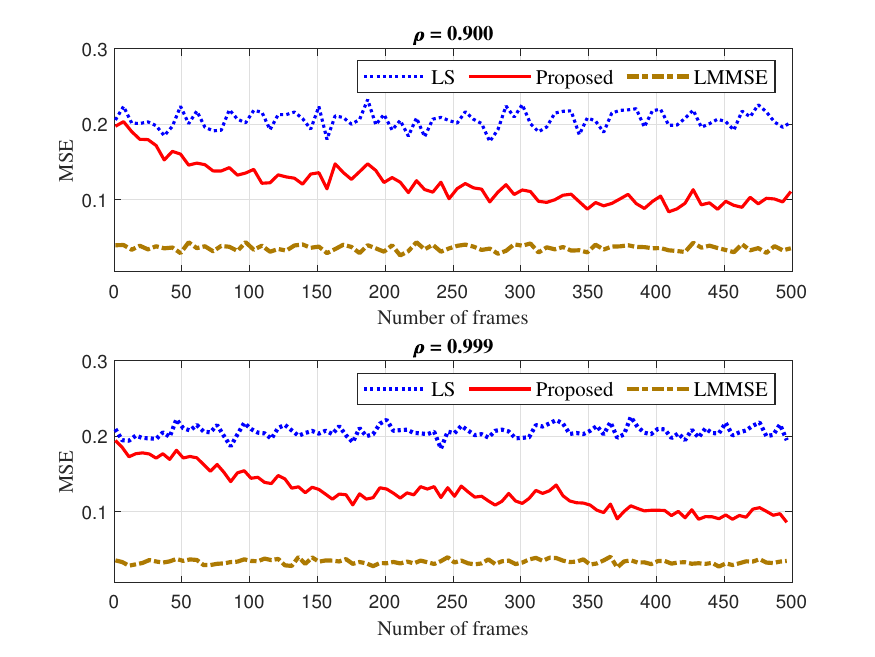}
        \caption{MSE comparison between our method, LS, and LMMSE vs. transmitted frames for correlated channels, under different correlation factors.}
        \label{fig:Correlation}
    \endminipage
\vspace{-5mm}
\end{figure*} 

\begin{figure*}[ht]
    \centering
    \minipage{0.32\textwidth}
        \includegraphics[width=\linewidth]{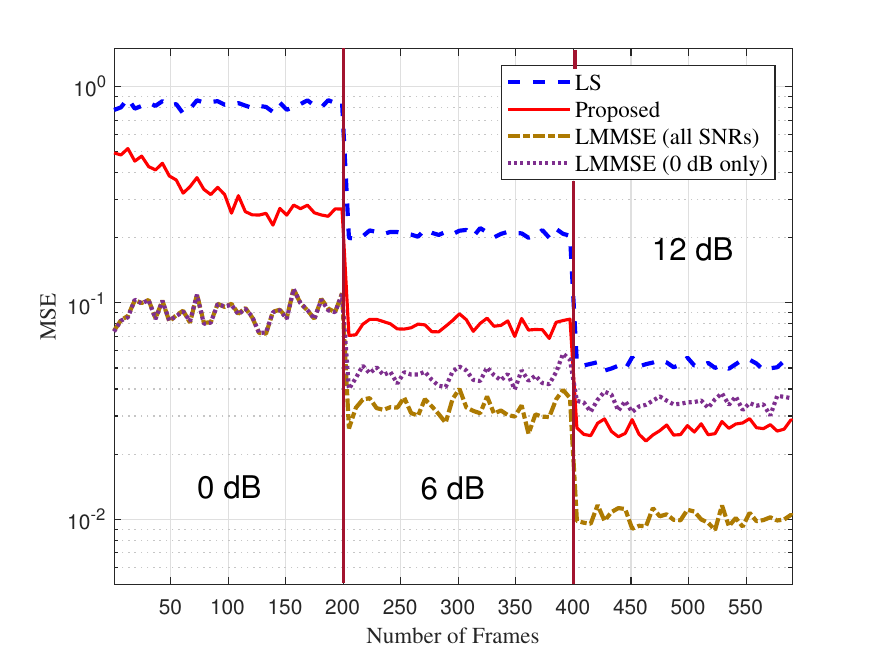}
        \caption{MSE comparison between our method, LS, imperfect LMMSE and LMMSE vs. the number of transmitted frames. Starting at 0 dB SNR, the SNR changes to 6 dB and 12 dB after transmitting 200 and 400 frames, respectively.}
        \label{fig:SNR}
    \endminipage\hfill
    \minipage{0.32\textwidth} 
        \includegraphics[width=\linewidth]{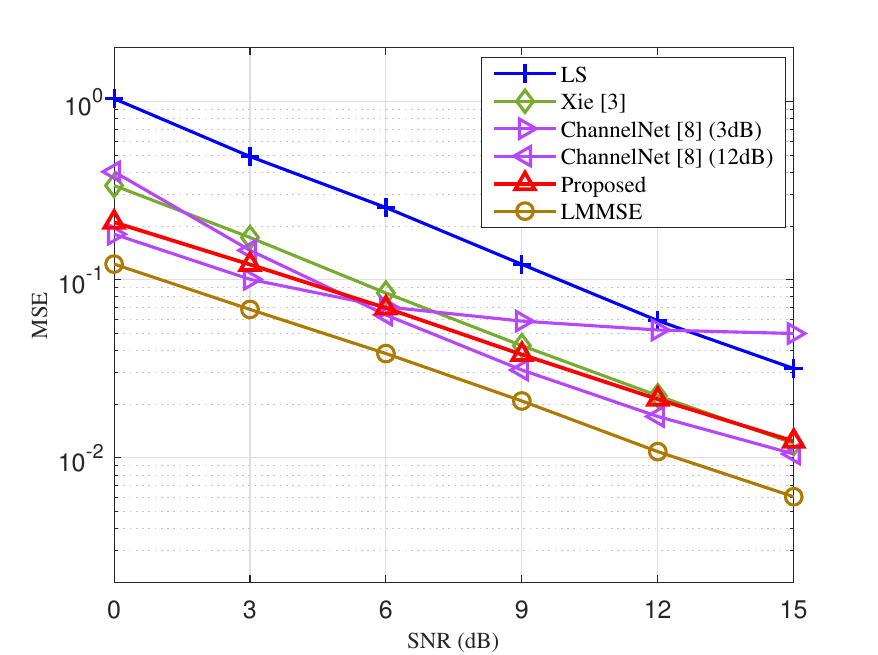}
        \caption{MSE comparison between our method, LS, the methods of~\cite{ce1} and~\cite{ML2}, and LMMSE vs. SNR.}
        \label{fig:MSE}
    \endminipage\hfill
    \minipage{0.32\textwidth}
        \includegraphics[width=\linewidth]{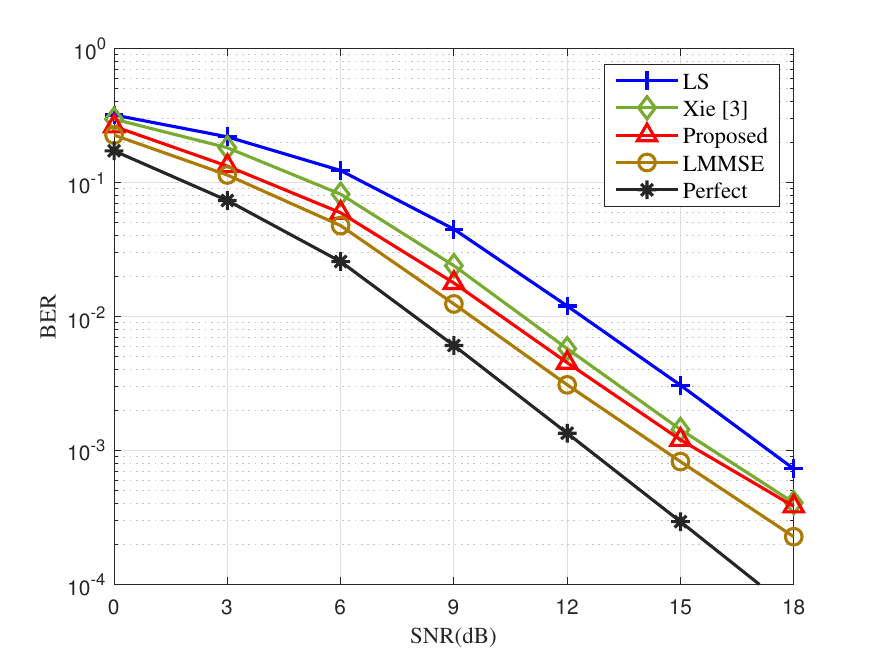}
        \caption{BER comparison between our method, LS, the method of~\cite{ce1}, and LMMSE vs. SNR.}
        \label{fig:BER}
    \endminipage
\vspace{-3mm} 
\end{figure*} 

We conduct a set of numerical experiments to analyze the performance of our proposed successive denoising method under different system settings.
We consider a MIMO OFDM system with parameters $N_t=4$, $N_r=4$, $K=32$, and $D=25$.
Unless stated otherwise, channels are generated from the exponential power delay profile (PDP) with $P=1$ and $L=8$.
We choose $\Delta=0.2$, $\alpha=0.3$, $\epsilon=0.5$, and $\gamma=1$, and measure MSE as follows:
\begin{equation}
    \text{MSE}=\frac{1}{N_tN_rK}\sum^{N_t-1}_{p=0}\sum^{N_r-1}_{q=0}\sum^{K-1}_{k=0}\big\vert \widehat{H}_{qp}^{(k)}-H_{qp}^{(k)}\big\vert ^2.
\end{equation}

We evaluate the learning performance of our method over a fixed set of channel realizations for different values of state dimension $M$ and channel length $L$ in Fig.~\ref{fig:Iter}.
For both channel lengths used, learning in both cases with $M=8$ result in lower MSE but take more iterations to converge.
This is because larger state dimensions generally require longer training times, but provide better performance by the end of the process.
We next consider channels with various time correlations and evaluate the performance of our method in Fig.~\ref{fig:Frames} (for i.i.d. channel generation) and Fig.~\ref{fig:Correlation} (for correlated channels).
The correlated channels are generated via Gauss-Markov process~\cite{Guass} with a correlation factor $\rho$.
As seen in Fig.~\ref{fig:Frames}, denoising over uncorrelated channels converges after 300 frames with a constant learning slope.  
Fig.~\ref{fig:Correlation} reveals that denoising over correlated channels exhibits a faster convergence (around 150 frames) due to stationarity of the channels.

To verify the robustness of our method against statistical variations of channels, MSE performance over time with varying SNR conditions is depicted in Fig.~\ref{fig:SNR}.
Starting at 0 dB SNR, the SNR changes to 6 dB and 12 dB after transmitting 200 and 400 frames, respectively.
An ideal LMMSE estimation case (i.e., SNR levels are always known) and an imperfect LMMSE estimation that only has the knowledge of initial channel statistics are considered.
The results demonstrate that compared to the degraded performance of LMMSE estimation with inaccurate channel knowledge, our method is able to keep its consistent performance relative to the ideal LMMSE estimation regardless of the channel condition.

Fig.~\ref{fig:MSE} depicts the MSE performance of our method over different SNRs.
We also include the results from the algorithms proposed in~\cite{ce1} and~\cite{ML2} for comparison.
From ChannelNet~\cite{ML2}, two curves each obtained from two different training datasets (3 dB and 12 dB SNRs) are included.
The results show that our method achieves an approximate 6 dB performance gain as compared to the LS estimation.
Our method outperforms the one in~\cite{ce1} especially in the low SNR regime, since the noise undetected by our proposed threshold becomes more dominant at high SNRs.
Both cases of ChannelNet~\cite{ML2} achieve lower MSE than our algorithm when SNR conditions are close to the level on which they were initially trained. 
Nevertheless, their performance  significantly degrades (e.g., see ChannelNet (3dB) evaluated at 12dB SNR) as the testing condition deviates from that of their training, which is the drawback of supervised learning methods. 
Our method, on the other hand, exhibits a consistent performance over all the SNRs, suggesting its generalizability.
This comes with the benefit of not relying on any training datasets, as well as without requiring any knowledge of operating SNR. 

Finally, we investigate bit-error rate (BER) performance of our method in Fig.~\ref{fig:BER}, where QPSK and an LDPC code of rate $R=\frac{1}{5}$~\cite{5G} with hard-decision decoding are used for data modulation and encoding/decoding, respectively.
Also, we used the baseline of~\cite{ce1} since it provides the closest performance to ours as compared to~\cite{ML2}.
The BER performance under perfect channel knowledge (i.e., when $H^{(k)}_{qp}$ is known at the receiver) is included to show the ideal performance.
The results verify that our algorithm achieves performance comparable to that of LMMSE estimation.

\section{Conclusion}\label{sec:conclusion}

We considered MIMO OFDM systems and proposed a novel channel estimation via successive denoising based on RL.
We proposed channel curvature as an effective metric to quantify channel estimation quality.
We derived the magnitude threshold of channel curvature to identify the target of denoising among subcarriers.
We then formulated the channel denoising procedure as an MDP and utilized a Q-learning approach to optimally decrease the MSE.
Through numerical results we showed that our method achieved a significant performance gain over the LS estimation and outperforms existing channel estimation techniques.
Our method does not require a prior knowledge on channel statistics, operating SNR, and a pre-labeled datasets for training, and hence dynamically adapts to variations in channel conditions.
These properties make our method practical in wireless systems with time varying channels where channel statistics are unknown. 

\section*{Acknowledgment}
 D. J. Love was supported in part by the National Science Foundation (NSF) under grants CNS1642982, CCF1816013, and EEC1941529. C. G. Brinton was supported in part by the NSF under grants AST2037864. T. Kim was supported in part by the NSF under grants CNS1955561.

\balance{
\bibliographystyle{IEEEtran}
\bibliography{IEEEfull,mybib}
}

\end{document}